\documentstyle[11pt,paspconf,psfig]{article}

\begin{document}

\title{Cosmic Rays in the Galactic Center Region}

\author{W. Rhode$^{1,2}$, T.A. En{\ss}lin$^3$, P.L. Biermann$^{3,4}$}
\affil{$^1$ Physics Dept., University of California,
    Berkeley, CA 94720, USA}
\affil{$^2$ Physics Dept., Univ. of Wuppertal, D-42097 Wuppertal, Germany}
\affil{$^3$ Max Planck Institute for Radioastronomy, D-53010 Bonn,
Germany}
\affil{$^4$ {\it Speaker at the conference}}

\begin{abstract}
EGRET data on the Gamma ray emission from the inner Galaxy have shown a rather
flat spectrum.   This spectrum extends to about 50 GeV in photon energy. 
It is usually assumed that these gamma-rays arise from the interactions
of cosmic ray nuclei with ambient matter.
Cosmic Ray particles have been observed up to $3 \, 10^{20}$ eV, with many
arguments suggesting, that up to about $3 \, 10^{18}$ eV they are of Galactic
origin. Cosmic ray particles get injected by their sources, presumably
supernova explosions.  Their injected spectrum is steepened by diffusive losses
from the Galaxy to yield the observed spectrum.  As cosmic ray particles roam
around in the Galactic disk, and finally depart, they encounter molecular
clouds and through p-p collisions produce gamma rays from pion decay.  The
flux and spectrum of these gamma rays is then a clear signature of cosmic rays
throughout the Galaxy.  Star formation activity peaks in the central region of
the Galaxy, around the Galactic Center, the focus of this meeting.  Looking
then at the gamma ray spectrum of the central region of our Galaxy yields clues
as to where the cosmic ray particles interact, and with what spectrum.  Using
the FLUKA Monte-Carlo, we have modelled this spectrum, and find a best fit for
a powerlaw spectrum of cosmic rays with a spectrum of 2.34, rather close to the
suggested injection spectrum for supernovae which explode into their own winds.
This suggests that most cosmic ray interaction happens near the sources of
injection; it has already been shown elsewhere that this is consistent with
the spectrum of cosmic ray nuclei derived from spallation.  One important 
consequence is that cosmic ray heating and ionization should be strong in the
Galactic Center region.

\end{abstract}


\section{Introduction}

The spectrum of Galactic cosmic ray particles extends to probably $3 \,
10^{18}$ eV; the various contributions have been reviewed extensively by
Wiebel-Sooth \& Biermann (1998), and the basic fits to the data for the
various chemical elements have also been given in Wiebel-Sooth, Biermann \&
Meyer (1998).  These cosmic ray particles interact with interstellar matter,
and so spallate to produce the secondary nuclei (see, e.g., Garcia-Munoz et al.
1987), as well as the gamma ray emission above GeV photon energy (see, e.g.,
Stecker 1971).

There is a new AGASA paper (Hayashida et al. 1998) suggesting that
neutrons near $10^{18}$ eV are seen coming from both the Galactic
Center region as well as the Cygnus region, detected by a correlation
in arrival direction.  The data suggest neutrons, because protons at
such an energy cannot get through the Galaxy on a straight line path
with its magnetic field, since at that energy the Larmor radius is
about 1 kpc.  The data cannot also easily be explained as gamma ray
photons, since then the correlation would be stronger at lower energy
even, where nothing is seen.  Of course, neutrons are produced by
isospin flip in p-p collisions.  It is worth remembering, that the
Galactic Center region, as well as the Cygnus region are the prime
candidate regions for star formation and supernova activity in the
Galaxy, as clearly shown in radio, far-infrared and gamma ray data.
Neutrons also can get here within their life time from the Galactic
Center, but not at much lower energy. This leaves neutrons as the most
probable origin of these events.  If this result is accepted with the
interpretation as neutrons, then it strongly supports the argument
that cosmic rays are indeed Galactic up to $10^{18}$ eV.  However, it
is also immediately obvious, that the flux of cosmic ray particles
required in the source regions to produce so many neutrons as
suggested by the AGASA data, needs to be rather high relative to the
flux observed at Earth.  This then leads to considerable ionization
and heating by low energy cosmic rays in the Galactic Center region.

In this paper we wish to review first the interpretations of the
cosmic ray particles in the Galactic Center region, and then show that
a simple concept may be sufficient to explain the new data from EGRET,
which show a rather flat gamma ray spectrum.  This is a severe test
for any theory of cosmic ray origin.

\section{Origin of high energy cosmic rays}

Our Galactic Center harbors a black hole, which probably went through many
activity episodes during its growth.  Therefore we want to ask first whether
this activity could possibly explain high energy cosmic rays, and as a
consequence gamma rays.

Biermann \& Strittmatter (1987) have shown that radio galaxy hot spots can
accelerate protons to about $10^{21}$ eV.  Scaling this result with the power
of the underlying source and using the jet/disk-symbiosis picture developed
by Falcke et al. (1995 and later papers) we obtain for the maximum proton energy

\begin{equation}
E_{p,max} \; = \; 6.7 \, 10^{20} \, Q_{jet, 46}^{1/2} \, {\rm eV}
\end{equation}

\noindent where $Q_{jet, 46}$ is the power of the jet in units of $10^{46}$
erg/s.  The most extreme inferred jet luminosity is about $3 \, 10^{47}$ erg/s,
and so energies up to

\begin{equation}
E_{p,max} \; = \; 4. \, 10^{21} \, {\rm eV}
\end{equation}

\noindent appear possible (Biermann 1998a).  Therefore radio galaxies and their
various counterparts such as compact radio quasars (see also Farrar \&
Biermann 1998) are clearly a suitable source for high energy cosmic
rays.  The jet-disk symbiosis does seem to work down to stellar size
black holes (Falcke \& Biermann 1998), and so we may be permitted to
use it for intermediate powers of a proposed source.

The jet power of our Galactic Center, assuming that the compact radio source
does signify the existence of a jet, is 

\begin{equation}
Q_{jet} \; = \; 5 \, 10^{38} \, {\rm erg/s}
\end{equation}

\noindent and so $E_{p, max} \, = \, 1.5 \, 10^{17}$ eV.  Therefore the
activity of our central black hole is insufficient to produce neutrons at
$10^{18}$ eV, and so is unlikely to help to explain the correlations
in arrival directions in the data.

The Galactic Center region does harbor many interesting binary systems, some of
which are referred to as mini-quasars; however, there again, their power is
just not sufficient to explain particles near $10^{18}$ eV.

There is a new hot disk model, where weakly relativistic protons produce
various secondaries in their interaction in the disk (Mahadevan 1998), but this
model also cannot explain any particles at $10^{18}$ eV.

Therefore we propose to explore in the following the activity and cosmic ray
injection properties of supernovae focussing on those supernovae that explode
into their own stellar winds (Biermann 1997).

\section{Galactic Cosmic Rays}

In a series of papers Biermann et al. (1993 and later) have proposed that
cosmic rays get injected from three sites predominantly:

\begin{itemize}
\item{}  Supernovae that explode into the interstellar medium.
\item{}  Supernovae that explode into their own stellar wind.
\item{}  Radio galaxies and compact radio quasars.
\end{itemize}

The predictions of these models have been given in various reviews, and we
summarize here briefly:

The cosmic ray particles which interact the most derive from the
wind-supernovae.  This happens since massive stars explode close to
their birthplace, where the original material is still around from
which they formed (Biermann \& Tinsley 1974).  Their source spectrum
has been predicted to be $E^{-2.33 -0.02 \pm 0.02}$ below the knee at
$5 \, 10^{15}$ eV particle energy.  For particles above the knee the
corresponding prediction is $E^{-2.74 - 0.07 \pm 0.07}$.  This
peculiar way of writing the expected theoretical error range signifies
an asymmetric error distribution, extending here from the most
probable value of 2.33 to 2.37 in the first case.  The bend (to
explain the knee) has been predicted to be at 600 $Z$ TeV, where $Z$
is the charge of the chemical element nucleus under consideration, and
the cutoff is near 100 $Z$ PeV.  Because the energy of the bend
depends on the charge, the element abundance gets heavier at the knee,
as noted already by Peters (1959, 1961).

To understand the concept of wind-supernovae we must remember that
massive stars come in a variety of clothes, which are their different
wind shells: First, stars above 8 solar masses, but below about 15
solar masses explode as supernovae, but do so directly into the
interstellar medium.  This is the classical case.  Then stars above 15
but below about 25 solar masses explode into their wind, and that wind
may be powerful enough to sweep up interstellar material into a shell
mixed with shocked wind material, but the shell is still rather thin.
The material in this shell is enriched in Helium from the nuclear
reactions inside the star.  Finally, above about 25 solar masses the
winds get very powerful, leading to Wolf-Rayet stars, and is heavily
enriched. In this case the wind-shell may be rather thick.

The transport through the Galaxy is described by a diffusion
coefficient, which depends on the wavefield that derives from a
Kolmogorov spectrum of interstellar turbulence, and so the cosmic ray
spectrum is steepened by 1/3 (see, e.g., Biermann 1995), to yield
$E^{-2.67 -0.02
\pm 0.02}$ below the knee.

This result can be directly compared with the data for Helium through Iron,
which give a best fit of $E^{-2.64 \pm 0.04}$ (Wiebel-Sooth, Biermann \& Meyer
1998).  

However, where the interaction really happens is not clear.  There are many
possible points of view on this question, but two conceptually simple notions 
are
documented in the literature:

First, there is the CR-standard model (see, e.g., Garcia-Munoz et al.,
1987) that the average cosmic rays interact with the interstellar
matter.  In such a picture the gamma rays should have a spectrum that
nicely fits the average cosmic ray spectrum, near $E^{-2.7}$.  In such
a picture the secondary to primary ratio of spallation products such
as boron derived from carbon spallation gives the spectrum of
interstellar irregularities with an implied energy dependence of the
leakage time scale as $E^{-0.6}$.  One problem with this argument is
that there is little evidence for such a spectrum of irregularities
(Biermann 1995), but it does give a good fit.

Second, there is the notion that most cosmic ray interaction happens near the
source (Biermann 1998b).  In such a picture the spallation leading to secondary
nuclei production happens in the shell around the stellar wind, when the 
supernova 
induced shock smashes through that shell.  This leads to an energy 
dependence for the local leakage time scale of $E^{-0.55}$ (Biermann 1998b);
however,  this happens only when the shell is thick enough to allow diffusive 
interaction to be dominant over convective losses.  This latter 
process is likely to dominate for the more abundant, but thinner 
shells around slightly lower mass stars.  Therefore, in such a picture 
we expect that the more common stars in the range 15 to 25 solar masses 
would produce most gamma rays.  And as a corollary we expect that the gamma rays
should correspond to the injection spectrum.

\subsection{The failure of the CR-standard model}

The standard model has been explored in two papers recently, using the best
EGRET data (Hunter et al. 1997, Mori 1997).  The standard model fails by a wide
margin.

The failure is due to the spectrum.  The observed gamma ray spectrum
is just too flat in order to be produced by a cosmic ray spectrum near
$E^{-2.7}$.

There are several ways out of this conundrum.

First, one might argue that the Monte-Carlo codes used to predict the gamma
rays are not good enough.  This is what Mori (1997) has tried.  The
uncertainties in the Monte-Carlos are not sufficient to explain the flat
spectrum.

Second, one might argue, that pion decay does not explain the data.
This has been tried by Pohl \& Esposito (1998).  They suggest that the
spectrum can be partially derived from inverse Compton scattering off
a population of energetic electrons produced by supernova remnants.
In the progressive leakage of the electrons as a function of energy
and time from injection the observed spectrum can be matched.  If the
new AGASA data are correctly interpreted with arising from energetic
neutrons, then cosmic ray nucleon interaction is about as high as can
possibly be, and so it is difficult to see how to avoid gamma ray
production from pion decay being a strong contributor.

Conversely, a success of an alternative model that also explains other data
would be very helpful, and this is what we have tried.

\subsection{A fit to the data}
\begin{figure}
\vspace{-1.5cm}
\centerline{\psfig{figure=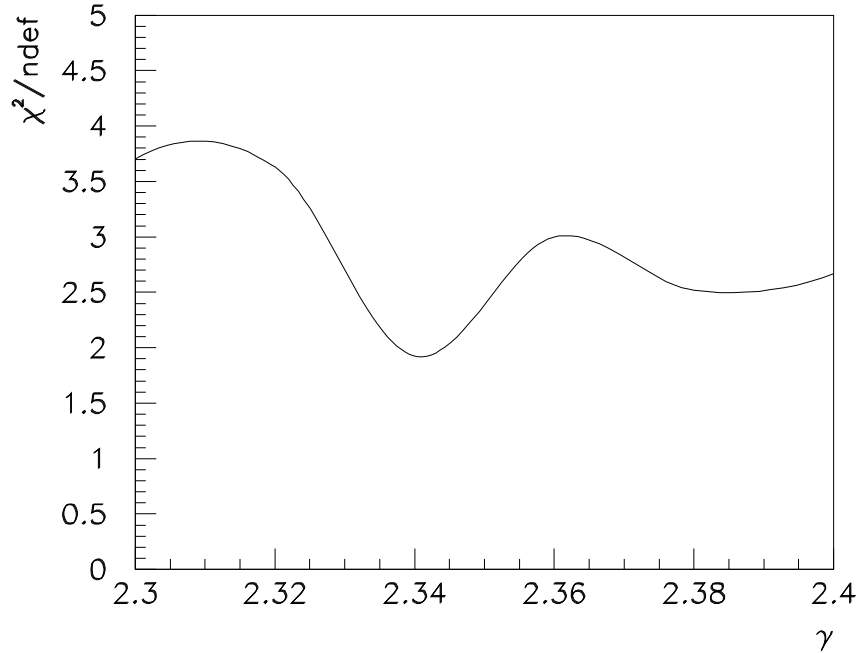,width=13cm}}
\vspace{-1cm}
\caption{$\chi^2_{red}$ versus spectral index.} 
\label{fig-1} \end{figure}
Therefore we have adopted a simple powerlaw model for the
Bremsstrahlung and inverse Compton contribution and then fitted the
data with one main parameters in mind: The power law spectrum of the
cosmic rays; this law is given as a strict power law in momentum from
many MeV to many GeV and beyond in energy.

We have tried three different Monte-Carlo codes from CERN and
Fermi-Lab to do this analysis, and we have adopted for this work the
code FLUKA (specifically the version GEANT3.21/FLUKA from the CERN
library). This code could readily be adapted to include the subtle
effects of Helium for instance.

Fig. 1 gives the $\chi^2_{red}$ fit as a function of spectral index.
The wiggle at spectral index 2.36 is due to a systematic pattern in the 
data from the primary data analysis.  In the plot shown here a clear 
minimum is visible at spectral index 2.34.
This minimum is at a level of $\chi^2_{red}$ of 1.9. Taking into account
the uncertainties of the IC and Bremsstrahlung contribution this is 
quite acceptable.
Fig. 2 gives the resulting fit, which still shows some model dependent waves
from the finite resolution of the Monte-Carlo used.  

\begin{figure}
\vspace{-1.5cm}
\centerline{\psfig{figure=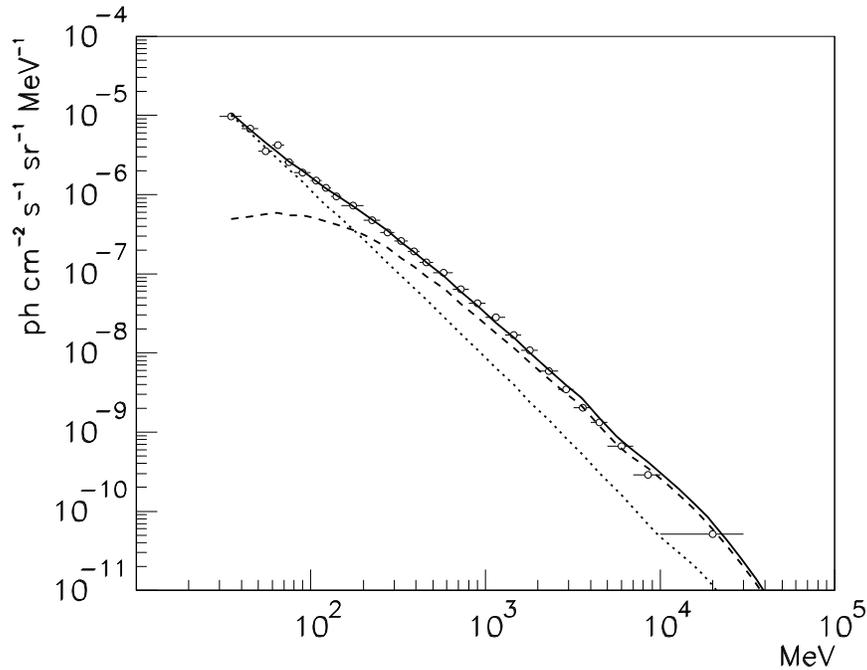,width=13cm}}
\vspace{-1cm}
\caption{The fit to the gamma ray data of the inner Galaxy.} \label{fig-2}
\end{figure}

The next step will be to check how high in energy we can push this
model of cosmic ray interaction near the source; there are severe
limits now on the inner Galaxy from the CASA-MIA experiment (Ong
1998).

\section{Consequences}

First of all, what the agreeable fit demonstrates is that there is a
spectrum for the cosmic rays that fits their gamma ray emission.  This
spectrum is consistent with a power law, and is in fact quite close to
the original prediction of a source spectrum for wind-supernovae.

Second, it shows that source related interaction may be worth pursuing
in detail.  What has not yet been done here, is a fit to the detailed
isotope abundances (see the recent discussion of this point by
Westphal et al. 1998).  Together with the paper presented at the
Hirschegg conference (Biermann 1998b) this means that there is a
viable proposal how to explain a) the cosmic ray spectrum itself, b)
the gamma ray spectrum, and c) the spectrum of spallation secondaries.

If this scenario could be confirmed there are some consequences also
for the stars with strong winds:

\begin{itemize}
\item{} Wolf-Rayet and OB stars have shock waves running through their winds.
\item{}  These shocks accelerate electrons and produce observed radio emission
(Biermann \& Cassinelli 1993).
\item{}  These shocks accelerate also protons, resulting in a steep pion decay
spectrum; this spectrum is steep because the Alfv{\'e}nic Machnumber of these 
shocks
is low. 
\item{}  These shocks also accelerate nuclei, which can give rise in
collisions  to spallation products in an excited nuclear state, then explaining
gamma ray lines (Nath \& Biermann 1994b) from active regions of star formation.
\end{itemize}

Finally, to summarize the essential idea again, there are consequences of this
scenario as well for exploding stars with strong winds:

\begin{itemize}
\item{}  The supernova shock races through the wind.
\item{}  The shock accelerates particles.
\item{}  Cosmic ray injection of elements such as helium and most
heavier elements originates from this acceleration.
\item{}  Once outside their site of origin the protons (and other nuclei) at
energies below about 50 MeV use up their ionization and heating power near their
origin (Nath \& Biermann 1994a).  
\item{}  In the Galactic Center region the cosmic
ray induced ionization and heating should be high.
\end{itemize}

\acknowledgments

We acknowledge numerous discussions on these matters with many friends and 
colleagues, especially John Bieging, Alina Donea, Tom Gaisser, Stan Hunter, Bob
Kinzer,  Norbert Langer, Karl  Mannheim, Sera Markoff, Jim Matthews, Hinrich 
Meyer,
Rene Ong, Buford Price, Ray Protheroe, J{\"o}rg Rachen, Eun-Suk Seo, Todor 
Stanev, 
and Dave Thompson.   WR would like to express his gratefulness for the warm
hospitality he has received at Berkeley in the group of Buford Price.

\end{document}